\documentclass[journal,comsoc]{IEEEtran}

\IEEEoverridecommandlockouts
\ifCLASSINFOpdf
\else
\fi
\usepackage{booktabs}
\usepackage{makecell}
\usepackage{graphicx}
\usepackage{subfigure}
\usepackage{color}
\usepackage[pagebackref=false,breaklinks=false,letterpaper=true,colorlinks,citecolor=blue,linkcolor=blue, anchorcolor=blue, bookmarks=true]{hyperref}
\usepackage{algorithm}
\usepackage{algorithmic}
\usepackage{pifont}

\newcommand{\etc}{\emph{etc}}

\usepackage{cite}

\usepackage{lettrine} 

\title{Spectrum Cognition: Semantic Situation for Next-Generation Spectrum Management}
\author{
Hao Zhang, \IEEEmembership{Member, IEEE}, 
Fuhui Zhou, \IEEEmembership{Senior Member, IEEE}, 
\\Qihui Wu, \IEEEmembership{Fellow, IEEE}, 
and Chau Yuen, \IEEEmembership{Fellow, IEEE}
\thanks{
This work was supported in part by the National Natural Science Foundation of China under Grant 62222107, in part by the Yangtze River Delta Science and Technology Innovation Community Joint Research (Basic Research) Project under Grant BK20244006, in part by the National Key Research and Development Project under Grant 2023YFB2904500, and in part by the Basic Research Projects of Stabilizing Support for Specialty Disciplines under Grant ILF240041A24. (\emph{Corresponding author: Fuhui Zhou}.)

H. Zhang and F. Zhou are with the  Artificial Intelligence, Q. Wu is with the College of Electronic and Information Engineering, Nanjing University of Aeronautics and Astronautics, Nanjing 211106 China. They are also with the Key Laboratory of Dynamic Cognitive System of Electromagnetic Spectrum Space (Nanjing University of Aeronautics and Astronautics) and with the Ministry of Industry and Information Technology, Nanjing, 211106, China (emails: haozhangcn@nuaa.edu.cn, zhoufuhui@ieee.org, wuquhui2014@sina.com)

C. Yuen is with the School of Electrical and Electronic Engineering, Nanyang Technological University, Singapore 639798. (email: chau.yuen@ntu.edu.sg)
}
}

\begin{document}
\bstctlcite{BSTcontrol}

\maketitle

\begin{abstract}

In response to the growing complexity and demands of future wireless communication networks, spectrum cognition has emerged as an essential technique for optimizing spectrum utilization in next-generation wireless networks. This article presents a comprehensive overview of spectrum cognition, underscoring its critical role in enhancing the efficiency and security of future wireless systems through the innovative perspective of \lq\lq data processing to signal analysis to semantic situation\rq\rq. Semantic situation, as the highest level of spectrum cognition, enables the extraction of meaningful information from raw spectrum data to provide intelligent support for network decisions. We formally define spectrum cognition, clearly distinguishing it from traditional spectrum sensing, and delve into the latest advancements in both traditional and intelligent spectrum cognition frameworks, addressing key challenges in spectrum cognition. Furthermore, we propose concrete technical solutions to address these challenges, highlighting the transformative potential of semantic situation in shaping next-generation wireless systems. Our findings not only contribute to the theoretical understanding of spectrum cognition but also offer practical insights for its implementation in real-world scenarios.

\end{abstract}


\section{Introduction} 
\label{sec:introduction}

\IEEEPARstart{S}{INCE} wireless communication networks advance towards the future sixth-generation (6G) systems, the exploration into the capabilities and challenges of 6G becomes increasingly crucial, as highlighted in \cite{saad2019vision,wang2023road,singh2025towards}. 
For civilian aspects, 6G networks will become even more complex, which are envisioned to support a higher frequency range including terahertz (THz) bands, and a myriad of emerging applications, including augmented reality (AR), virtual reality (VR), and holographic communications. 
For military domains, 6G networks will not only become complex but also become more dynamic, mainly due to the rapid changes in the battlefield due to communication countermeasure. 
Moreover, with the integration of 6G and artificial intelligence (AI), the security challenges will be further exacerbated, since AI-based models are more vulnerable to adversarial attacks. 
These demands underscore the necessity for efficient spectrum utilization in 6G systems, a critical strategy to mitigate spectrum resource scarcity and prevent the inefficient use of spectrum resources. 
Therefore, spectrum cognition is imperative for the development and optimization of future wireless communication systems.

Traditional spectrum analysis, comprising spectrum sensing for identifying unused frequency bands and abnormal signal detection for identifying security threats, is increasingly inadequate for future wireless communications \cite{wu2024one}. Spectrum cognition is collecting, analyzing, and interpreting spectrum data to derive semantic meaning beyond mere signal detection. Unlike traditional approaches focused on binary “occupied/unoccupied” determinations, spectrum cognition encompasses the entire chain from data acquisition to semantic interpretation, with “semantic situation” as its core concept that analyzes context, spatio-temporal correlations, and evolutionary trends of spectrum data.

The significance of spectrum cognition for future wireless communications is multifaceted, offering intelligent spectrum analysis that identifies underutilized portions for improved efficiency, providing user-perspective insights that enhance experience, and strengthening network security against malicious users and AI adversarial attacks. By progressing from data processing through signal analysis to semantic situation understanding, spectrum cognition enables a more efficient, secure, and user-centric network, making it a promising technique for addressing the complex demands of next-generation wireless communications.

\begin{figure*}[t]
  \centering
  \includegraphics[width=0.95\linewidth]{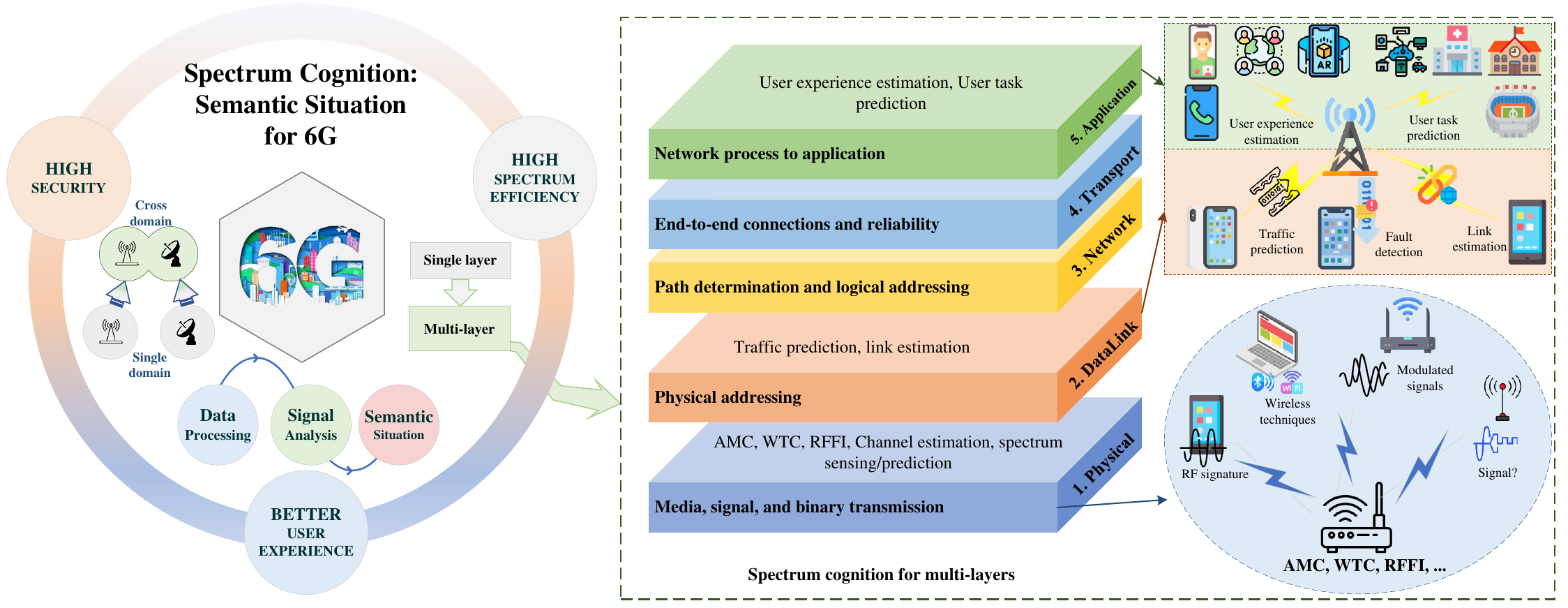}
  \caption{Spectrum cognition. The left part presents benefits for 6G, and the right part illustrates spectrum cognition for multi-layer applications such as automatic modulation classification (AMC), wireless technology classification (WTC), radio frequency fingerprinting (RFFI), traffic prediction, and user experience prediction.}
  \label{fig:vision}
\end{figure*}

Spectrum cognition represents significant advancements in the field of wireless communications, transcending the traditional boundaries of spectrum sensing, representing leaps in \lq\lq from data processing to semantic analysis to behavior reasoning", \lq\lq single domain to cross-domain", and \lq\lq single layer to multi-layer". 
\begin{itemize}
  \item \emph{\textbf{From data processing to signal analysis to semantic situation}}: Traditional spectrum analysis primarily focuses on data collection, such as identifying unused frequency bands. In contrast, spectrum cognition interprets this data to extract meaningful semantic information and to further analyze how these patterns translate into user and network behavior. By predicting future usage patterns and behaviors, network operators can proactively manage resources, leading to enhanced efficiency and user experience. 
  
  \item \emph{\textbf{From single domain to cross-domain}}: Traditional spectrum analysis operates within isolated domains such as frequency bands (e.g., sub-6GHz, mmWave), network types (e.g., cellular, Wi-Fi, satellite), or application areas (e.g., communication, radar, sensing). In contrast, spectrum cognition enables cross-domain analysis that integrates information across multiple dimensions. For example, it can simultaneously analyze cellular signals in sub-6GHz bands, Wi-Fi activity in unlicensed spectrum, and radar operations in mmWave frequencies to understand overall spectrum utilization patterns. Such comprehensive analysis is essential for 6G networks that must efficiently utilize diverse frequency ranges while supporting coexistence between communication, sensing, and positioning services.

  \item \emph{\textbf{From single layer to multi-layer}}: These advanced methods also expand the analysis from a single-layer perspective, such as the physical layer of network architecture, to a multi-layered approach. This includes considering factors from the network, transport, and application layers, thereby offering a multi-dimensional view of spectrum utilization. Such an approach is critical in addressing the complex and layered nature of modern wireless networks.
\end{itemize}

This paper aims to explore the intricacies of spectrum cognition, illustrating their impact on the future of wireless communications. 
The remainder of this article is organized as follows. Section \ref{sec_sota} introduces the state of the art of spectrum cognition. Following that, we present the technical details for the implementation of spectrum cognition in Section \ref{sec:key_technologies}. Then, we focus on the challenges and corresponding solutions proposed in Section \ref{sec_open_issues}. Finally, Section \ref{sec_conclusion} concludes this article.

\section{Representative Traditional and Intelligent Spectrum Cognition Frameworks}\label{sec_sota}

This section reviews the evolution of spectrum cognition approaches from traditional statistical methods to modern intelligent frameworks. We examine the strengths and limitations of detection-based, likelihood-based, and feature-based traditional methods, followed by machine learning and deep learning approaches that enable more sophisticated spectrum understanding.

\begin{figure}[t]
  \centering
  \includegraphics[width=0.95\linewidth]{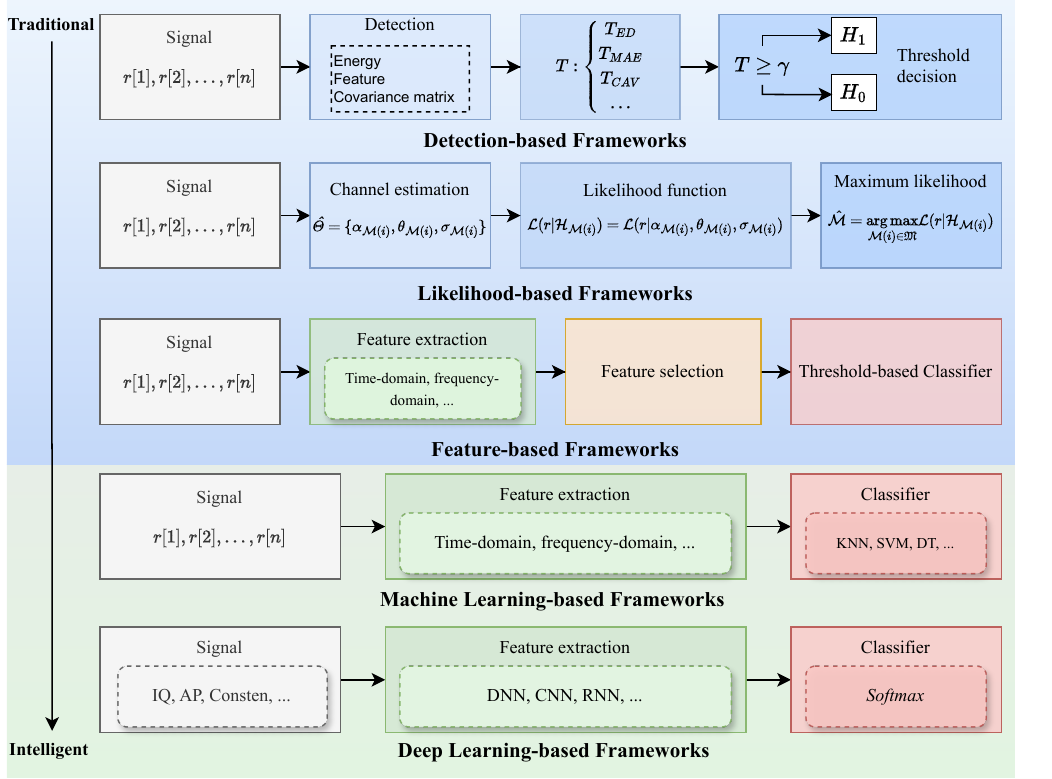}
  \caption{The development of spectrum cognition frameworks. Traditional spectrum cognition frameworks include detection-based, likelihood-based, and feature-based. Intelligent spectrum cognition frameworks include machine learning-based and deep learning-based.}
  \label{fig_methods}
\end{figure}

\subsection{Traditional Spectrum Cognition Frameworks}
Traditional spectrum cognition frameworks include decision-based (DB), likelihood-based (LB) and feature-based (FB) frameworks, as shown in Fig. \ref{fig_methods}.

\subsubsection{Detection-based spectrum cognition frameworks}
Detection-based (DB) frameworks evaluate detection statistics for each hypothesis using observed signal samples, followed by comparison for classification decisions. While fundamental and computationally efficient, DB methods offer limited accuracy in complex signal environments.

\subsubsection{Likelihood-based spectrum cognition frameworks}
Likelihood-based (LB) frameworks determine signal types by calculating probability distributions of observed samples. Despite their popularity, LB frameworks struggle with noise sensitivity and high computational complexity, presenting significant challenges for 6G systems.

\subsubsection{Feature-based spectrum cognition frameworks} 
Feature-based (FB) frameworks extract specific time-domain, frequency-domain, or other domain features using transformation and statistical techniques before classification. FB methods reduce computational complexity compared to LB approaches but face challenges in feature selection and classifier design, limiting their application to specific modulation types under specific scenarios.

\subsection{Intelligent Spectrum Cognition Frameworks}
Recent intelligent technologies offer enhanced capabilities for spectrum cognition but face implementation challenges in practical systems.

\subsubsection{Learning-Based Spectrum Cognition Frameworks} 

Fig. \ref{fig_methods} shows the frameworks of both machine learning-based and deep learning-based methods for spectrum cognition. Both ML-based and DL-based frameworks share a similar three-part structure: signal processing, feature extraction, and classification.

In ML-based frameworks, the received signals are first processed by transforming them into a specific format, followed by feature extraction using filter bank (FB) methods to extract pre-defined features. Finally, classifiers such as support vector machines (SVMs), K-nearest neighbors (KNNs), or decision trees are employed to classify the extracted features. However, ML-based frameworks fundamentally rely on FB frameworks, which inherently limits their performance to that of the underlying FB methods. Consequently, these frameworks share similar shortcomings with FB approaches, including challenges in proper feature selection and classifier design. Additionally, ML-based frameworks require substantial training data and face the risk of overfitting, which limits their ability to generalize to new, unseen data. The training process is also time-consuming.

DL-based frameworks follow a similar structure but with enhanced capabilities. Signals can be processed into various representations including image-related, sequence-related, and other data formats. These representations are then fed into deep learning-based feature extraction models, typically deep neural networks (DNNs), convolutional neural networks (CNNs), or recurrent neural networks (RNNs). The learned features are subsequently classified using \emph{softmax}-based classifiers, implemented as fully connected layers with \emph{softmax} activation functions.

While DL-based frameworks can achieve superior performance compared to traditional ML approaches, they present their own challenges. These models typically contain a vast number of parameters and require very large datasets for effective training. The training process demands substantial computational resources and time, which may not be readily available in practical systems. Furthermore, the internal workings of DL models are inherently complex and nonlinear, making them function as \lq\lq black boxes" that are difficult to interpret and debug, potentially limiting their applicability in scenarios requiring model transparency and explainability.

\subsubsection{Integrated Model-Based and Learning-Based Approaches}
Integrated Model-Based and Learning-Based Approaches combine strengths of both paradigms to address individual limitations. Physics-informed neural networks (PINNs) \cite{kokolakis2025safe} incorporate physical laws into neural architectures, while model-guided learning approaches use theoretical models to inform feature selection or network design. These integrated methods demonstrate superior performance with limited training data, enhanced interpretability, and reduced computational requirements compared to pure deep learning solutions.

\section{Technical Details for Implementation} 
\label{sec:key_technologies}

This section presents the technical architecture for implementing spectrum cognition through our proposed \lq\lq data processing-signal analysis-semantic situation" framework. We detail the progression from raw signal processing to intelligent semantic understanding, providing practical guidance for each implementation layer.

\begin{figure*}[t]
    \centering
    \includegraphics[width=0.9\linewidth]{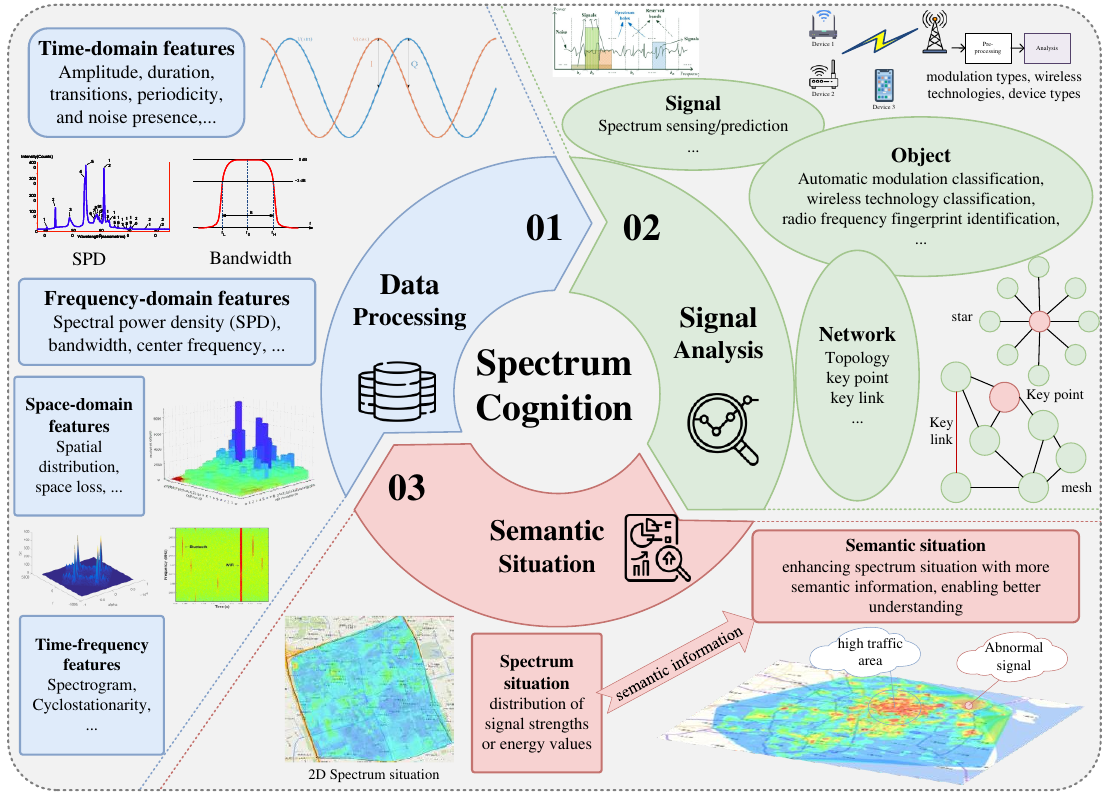}
    \caption{Technical details for implementation from the view of "data processing-signal analysis-semantic situation". The semantic situation component is highlighted as the core and highest level of spectrum cognition.}
    \label{fig_tech}
\end{figure*}

\subsection{Data Processing}
The received signals are typically represented in a raw data form, such as IQ samples, which is hard to use directly for signal recognition. Thus, the signal representation is a crucial step for spectrum cognition \cite{jiang2020advanced}. Signals can be represented in different forms, such as time-domain, frequency-domain, space-domain features and time-frequency domain \cite{zhang2025revolution}. 

\subsubsection{Time-domain features}
Time-domain features characterize signal properties that vary over time. Time-domain features include amplitude, propagation delay, latency, duty cycle, symbol rate, packet duration, synchronization Interval, coherence time, multipath fade duration, \etc. These time-domain features are essential for signal analysis, modulation recognition, and other aspects of wireless communication system design and optimization.

\subsubsection{Frequency-domain features} 
Frequency-domain features include bandwidth, carrier frequency, and spectral characteristics of the signal. These features provide valuable information about the spectral content and frequency-selective fading effects of wireless channels. They are useful for modulation classification, channel equalization, motion sensing, and other applications. 

\subsubsection{Space-domain features} 
Space-domain features include spatial correlation, spatial signatures, angle of arrival, space-time channel response, spatial diversity, radiation pattern, space loss, spatial fading statistics, multiple-input and multiple-output (MIMO) capacities, beamforming, \etc. Space-domain analysis is important for multi-antenna and multi-user wireless systems, providing spatial diversity and multiplexing gains. It allows the exploitation of spatial dimensions to improve wireless capacity and resilience.

\subsubsection{Time-frequency features}
Time-frequency features of a wireless signal refer to the characteristics of the signal in both the time and frequency domains. These features provide valuable information about the signal's behavior, modulation, and spectral content. Commonly used time-frequency features of wireless signals include power spectral density, spectrogram, instantaneous frequency, time-frequency distribution, and modulation recognition features. These time-frequency features are commonly employed in signal processing and communication systems to analyze and characterize wireless signals, enabling tasks such as signal classification, interference detection, spectrum sensing, and channel estimation.

\subsubsection{Other features}
Apart from the above four domain features, human-expert features include instantaneous time features, statistical features, and transform features are usually used for feature-based signal recognition approaches. Instantaneous time features consist of a series of parameters, namely, carrier amplitude, phase, and frequency. Statistical features include moments, cumulants, and cyclostationarity. Transformed features include Fourier transform, Wavelet transform, and $\mathcal{S}$ transform. These features provide powerful multi-domain perspectives on signal characteristics useful for wireless communications applications. 

\subsection{Signal Analysis}

\subsubsection{Signal-domain} Signal recognition in the signal domain includes spectrum sensing and spectrum prediction. 
Spectrum sensing involves scanning the radio spectrum, detecting occupied and unoccupied channels, and estimating signal levels. This allows radios to gain awareness of the spectrum environment.
Common sensing techniques include energy detection, matched filter detection, cyclostationary feature detection, radio identification, and \etc. Cooperative and distributed sensing can improve performance. 
Spectrum prediction utilizes the knowledge gained from spectrum sensing to anticipate and predict future patterns of spectrum occupancy and availability.
Prediction can be done using models like hidden Markov models, neural networks, Kalman filters, and \etc. Historical data is used to train the models.
Accurate sensing and prediction provide efficiencies in spectrum utilization compared to static spectrum allocation policies. It allows adaptation to real-time spectrum use.

\subsubsection{Object-domain} Signal recognition in the object domain is a process of identifying and classifying objects in a wireless signal. It involves extracting features from the signal and using them to train a classifier to recognize the object. 
Automatic modulation classification, wireless technology classification, wireless interference identification, and radio-frequency fingerprint identification are typical examples of signal recognition in the object domain. 

\paragraph{Automatic modulation classification}
AMC automatically identifies the modulation scheme of received signals. It's worth noting that the choice of features, feature selection methods, and classification algorithms can vary depending on the specific application, signal characteristics, and computational constraints. Additionally, the performance of AMC systems can be influenced by factors such as signal-to-noise ratio (SNR), channel impairments, and the presence of interference. Therefore, it's important to evaluate and optimize the AMC system's performance under various conditions.

\paragraph{Wireless technology classification}
The unlicensed 2.4 GHz industrial, scientific and medical (ISM) band can be freely shared by different wireless technologies such as Wi-Fi, ZigBee, and Bluetooth. These technologies must compete for the limited spectrum resources in this band while attempting to coexist harmoniously. With exposure to varying interference levels among the competing technologies, successful coexistence without disruption is critical for maintaining adequate wireless communication performance. Thus, wireless technology classification is adopted to classify the wireless technologies \cite{yuan2021multiscale}. 

\paragraph{Wireless interference identification}
WII refers to recognizing types of interference signals without any prior information. It is a promising technology in non-cooperative communication scenarios. The WII methods have various applications in military and civilian fields, including spectrum monitoring or intercepted signal recovery.

\paragraph{Radio-frequency fingerprinting identification}
RFF identifies devices based on their unique radio frequency characteristics. Every device emits unique radio frequency (RF) signals determined by its specific hardware construction and materials. RFF involves capturing these device-specific RF signals using wideband antennas and high-frequency receivers and analyzing the signals through various mathematical techniques to extract identifiable device fingerprints.

\subsubsection{Network-domain} 
Signal recognition in the network domain is a process of identifying and classifying the topology, key points, and key links in a wireless network. It involves extracting features from the signal and using them to train a classifier to recognize the network characteristics. Network topology identification, key point identification, and key link identification are the main applications of signal recognition in the network domain.

\paragraph{Topology identification}
Network topology identification refers to the process of determining the topological structure of a computer network, including the layout of nodes and connections. Identifying the network topology helps in analyzing its performance, vulnerabilities, and capabilities. 
The main aspects of network topology identification include determining the physical and logical layout of the network, identifying the network typology, finding the relationships between nodes, discovering routing and switching equipment, mapping the geographical location of nodes, and detecting dynamic changes.
Network topology identification involves combining automated scanning tools with manual verification to create an accurate map of the network's physical and logical structure. This provides visibility into the network's organization and highlights areas for optimization or security.

\paragraph{Key point identification}
Key point identification in a wireless network involves the process of identifying critical elements within the network infrastructure. These key points play a significant role in the network's functionality, performance, and security. Identifying and understanding these key points is crucial for effective network management, troubleshooting, and optimization. Some key points to consider in wireless network identification include access points (APs), routers and switches, network controllers, wireless LAN controllers, network security appliances, network monitoring tools, network segments and subnets, and \etc. 
Identifying key points in a wireless network provides network administrators with a comprehensive understanding of the network's structure, functionality, and critical components. It enables efficient network management, troubleshooting, performance optimization, and implementation of appropriate security measures.

\paragraph{Key link identification}
Key link identification in a wireless network involves the process of identifying and understanding the critical links that play a significant role in the network's functionality, performance, and connectivity. These key links, also known as important connections or pathways, are essential for establishing reliable communication and ensuring seamless operation within the wireless network. Here is a summary of key link identification in wireless networks: backbone links, access point links, point-to-point links, mesh links, redundant links, interference-free links, cross-connect links, \etc. By identifying these key links in a wireless network, network administrators can gain insights into the network's structure, performance, and potential areas of improvement. This understanding enables efficient network management, troubleshooting, capacity planning, and optimization of critical network pathways.

\begin{figure*}[!ht]
  \centering
  \includegraphics[width=0.9\linewidth]{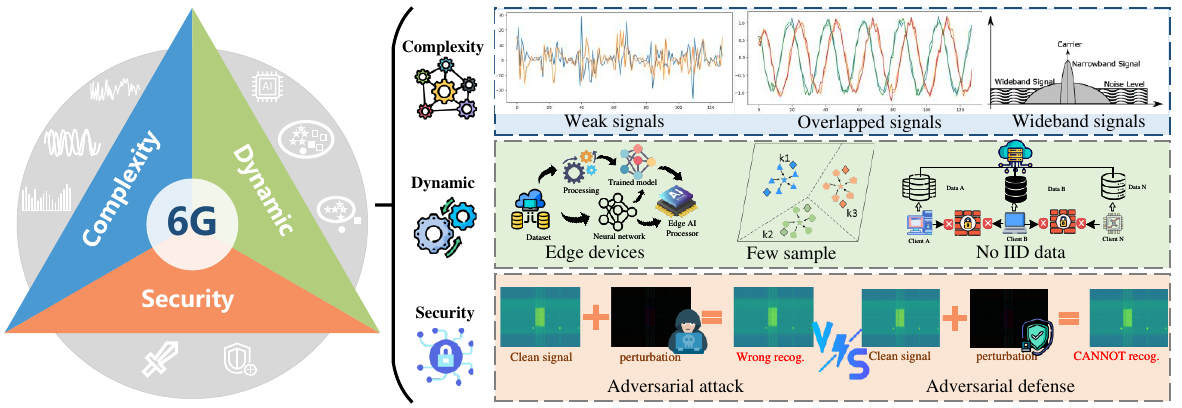}
  \caption{Open issues and challenges under the complex, dynamic, and open wireless networks.}
  \label{fig_challenges}
\end{figure*}

\begin{figure}
  \centering
  \subfigure[]{\includegraphics[width=0.9\linewidth]{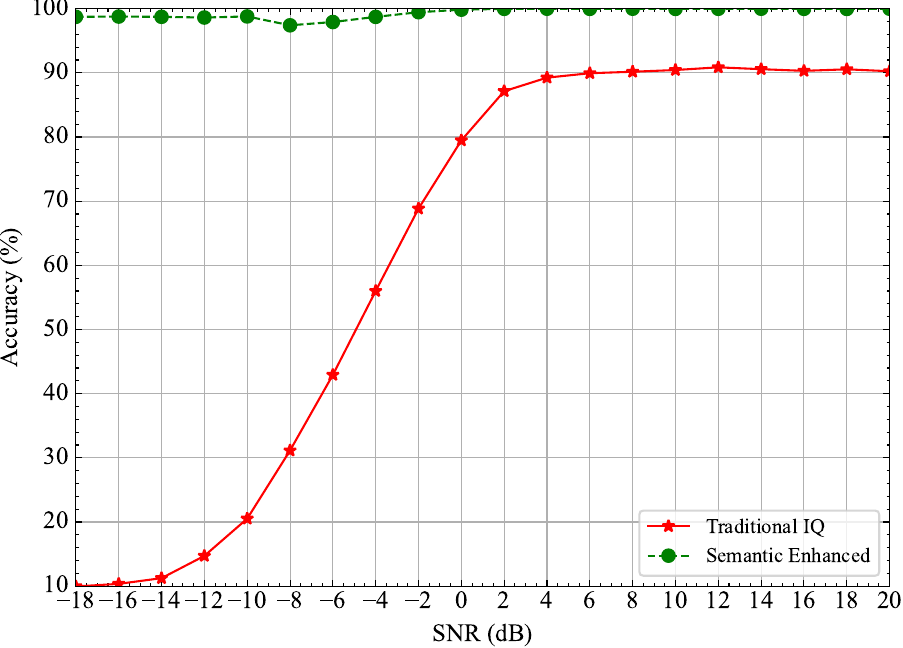}}
  \subfigure[]{\includegraphics[width=0.9\linewidth]{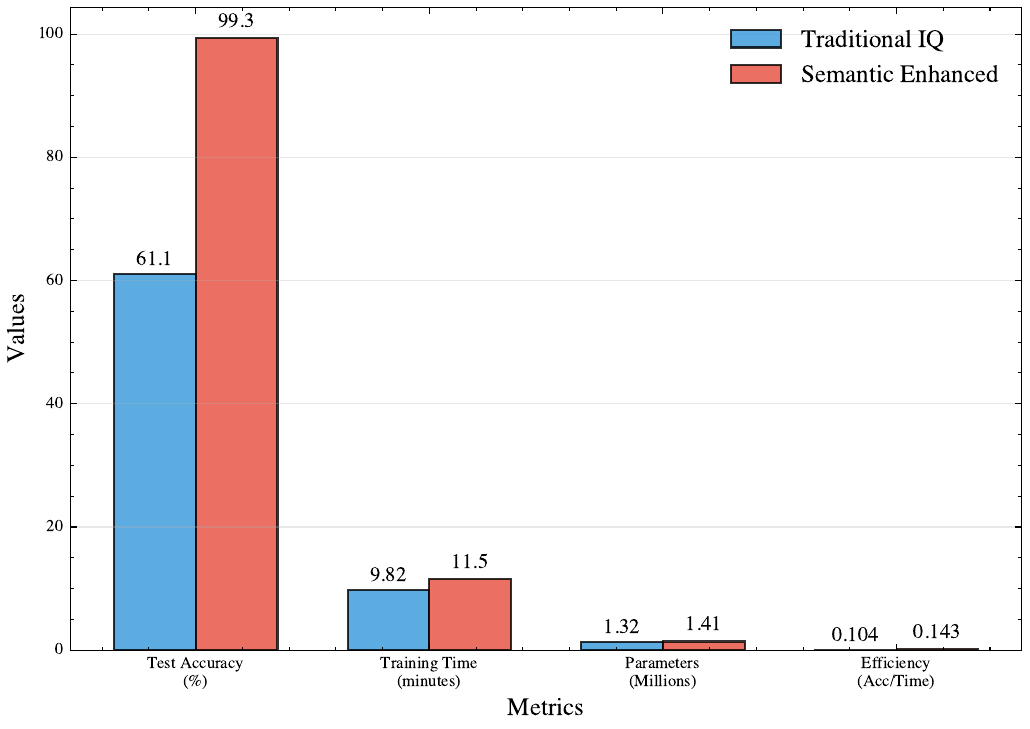}}
  \DeclareGraphicsExtensions.
  \caption{Performance comparison of the traditional IQ classifier and the semantic enhanced classifier, (a) accuracy over SNRs, and (b) the overall accuracy, training time, parameters, and efficiency.}
  \label{fig:performance}
\end{figure}

\begin{figure*}[!ht]
  \centering
  \includegraphics[width=0.85\linewidth]{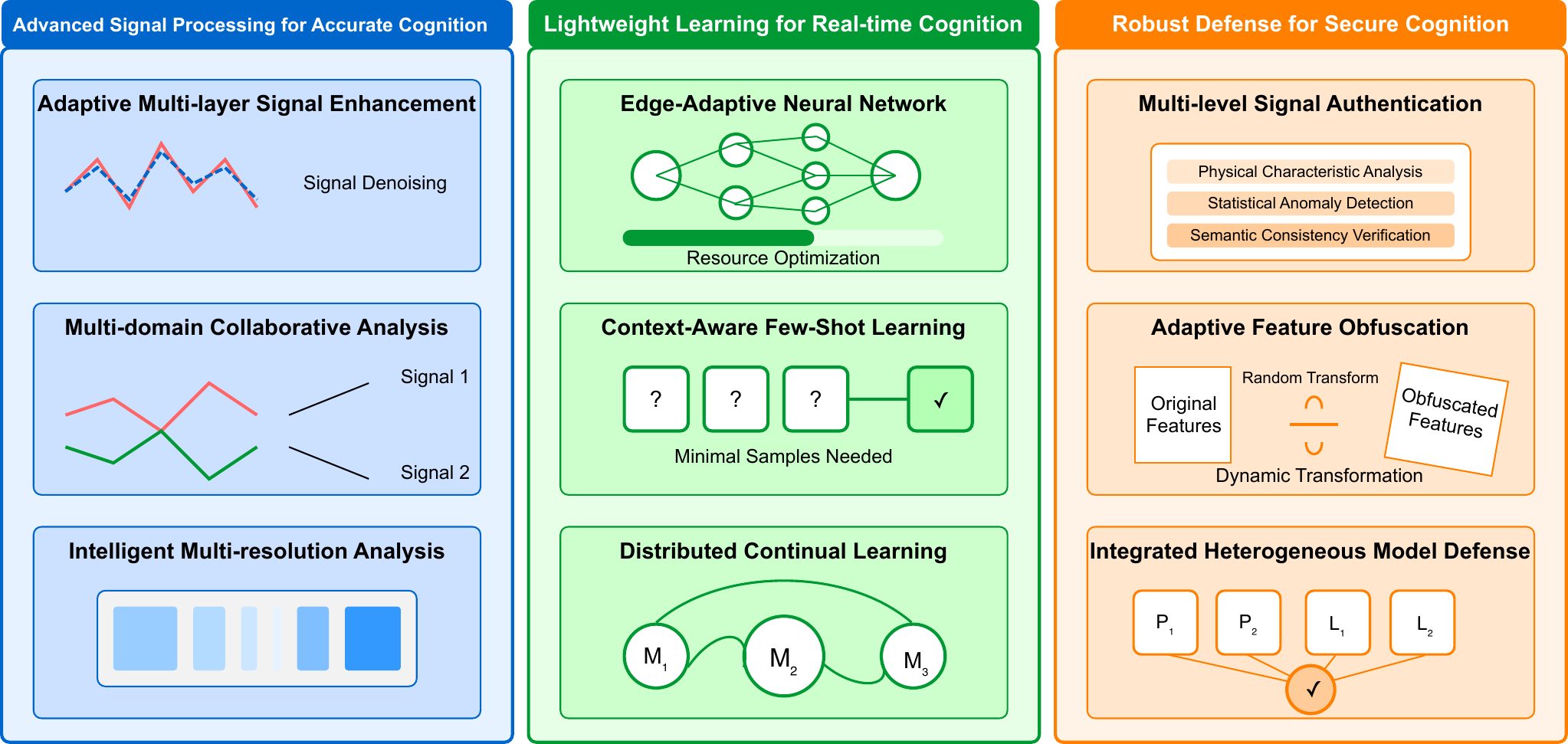}
  \caption{Possible solutions for the open issues and challenges.}
  \label{fig_solutions}
\end{figure*}

\subsection{Semantic Situation}

\subsubsection{Spectrum situation}
The spectrum situation is a visual representation of the radio frequency spectrum, showing the distribution of signal power across different frequency bands. It provides a comprehensive view of the spectrum environment, including occupied and unoccupied channels, signal strength, interference sources, and other relevant information. Spectrum situation is essential for spectrum management, interference detection, and spectrum sharing. 

\subsubsection{Semantic situation}
Semantic situation provides contextual understanding of spectrum activities and their underlying purposes. 
It can be structured into three hierarchical levels of semantics, 1) basic semantics, 2) relational semantics, and 3) intentional semantics. The basic semantics focus on identifying signal types, modulation schemes, and device characteristics. The relational semantics analyze interactions between devices, network topologies, and communication patterns. The intentional semantics infer communication purposes, predict future behavior, and understand user requirements. This hierarchical structure enables networks to understand spectrum activities from multiple dimensions, supporting more intelligent decision-making processes.

The formation of semantic situation is a progressive process that builds upon data processing and signal analysis to extract meaningful information, 1) First, data processing extracts signal features from raw spectrum data, 2) Next, signal analysis identifies objects and network structures, and 3) Finally, semantic analysis correlates these results to extract practically meaningful information. 
For example, by analyzing the movement patterns of multiple mobile devices identified through signal analysis, a semantic situation system can infer the occurrence of a large-scale event and predict resulting changes in spectrum demands. This level of understanding goes far beyond traditional spectrum sensing approaches. 
By incorporating semantic information (such as the state of the spectrum utilization, the topological relation among nodes, \etc) into the spectrum situation, we can achieve a higher-level understanding of the spectrum environment. The semantic situation involves analyzing the spectrum data to extract meaningful insights and identify patterns by using data processing and signal analysis. 
It provides a holistic view of the spectrum environment, enabling network operators to optimize spectrum utilization, detect anomalies, and improve network performance.

\subsection{Experimental Validation} \label{ssec:experimental_validation}
To validate the effectiveness of our proposed semantic situation framework, we conducted comprehensive experiments using the RML2016.10a \cite{zhang2024sswsrnet} dataset, comparing traditional IQ signal classification with semantic-enhanced approaches. Our experimental evaluation compared two classification methods: (1) a traditional IQ Classifier using only raw IQ signal data, and (2) a semantic Enhanced Classifier incorporating our proposed three-level semantic information including device type, scenario, and communication purpose. The experiments utilized the RML2016.04c dataset containing 11 modulation types (8PSK, AM-DSB, AM-SSB, BPSK, CPFSK, GFSK, PAM4, QAM16, QAM64, QPSK, WBFM) across an SNR range from -20dB to +18dB. We augmented the dataset with three semantic dimensions, device type encompassing 16 categories such as cellular\_base, wifi\_device, iot\_sensor, and satellite; scenario covering 12 categories including urban, rural, indoor, and emergency; and communication purpose comprising 11 categories like data, voice, control, and broadcast. Both classifiers employed identical CNN architectures for IQ processing, with the semantic model incorporating additional embedding layers for semantic features and a fusion mechanism to integrate the multi-level semantic information.

As shown in Figure \ref{fig:performance}, The experimental results demonstrate the significant effectiveness of our proposed semantic situation framework. The semantic enhanced Classifier achieved a remarkable test accuracy of 99.35\%, representing a substantial improvement of 38.3 percentage points over the Traditional IQ Classifier's 61.05\% accuracy. This dramatic performance gain validates that incorporating three-level semantic information (device type, scenario, and communication purpose) provides crucial contextual understanding that greatly enhances signal classification capabilities. The improvement comes with minimal computational overhead, as the semantic model required only 17.4\% more training time (692.5s vs 589.3s) and 6.3\% additional parameters (1.41M vs 1.32M), demonstrating the efficiency of our approach. These results clearly indicate that semantic context serves as a powerful complement to traditional IQ signal processing, enabling more robust and accurate modulation recognition across diverse scenarios.

\section{Challenges and Solutions for Spectrum Cognition in 6G Networks} 
\label{sec_open_issues}

This section discusses the open issues and challenges in spectrum cognition for 6G networks, focusing on three key aspects: complex electromagnetic environments, real-time processing requirements, and security vulnerabilities in open wireless networks. For each challenge, we propose innovative solutions that leverage advanced signal processing and machine learning techniques to enhance spectrum cognition capabilities.

\subsection{Accuracy Challenges in Complex Electromagnetic Environments}

Complex electromagnetic environments in 6G networks, characterized by low SNRs, overlapping signals, and ultra-wideband signal diversity, pose significant challenges for accurate semantic situation analysis. Traditional signal processing struggles to separate and interpret signals in such conditions, leading to degraded performance in spectrum cognition tasks like modulation classification and interference detection.

To address the challenges posed by complex electromagnetic environments, this part presents innovative signal processing techniques that form the foundation for accurate semantic situation analysis. The possible solutions include adaptive multi-layer signal enhancement methods that combine compressive sensing with deep denoising networks to improve signal detection in low SNR environments \cite{su2023signal}; multi-domain collaborative analysis algorithms that perform simultaneous analysis across time, frequency, and spatial domains to achieve high separation accuracy in overlapped signal environments; and intelligent multi-resolution analysis systems for 6G ultra-wideband signals that dynamically adjusts sampling rates and processing windows based on signal characteristics, substantially reducing processing burden while maintaining high accuracy for real-time semantic analysis of complex signals.

\subsection{Real-Time Processing Requirements in Dynamic Environments}

The dynamic nature of 6G networks, driven by rapid changes in the electromagnetic environment, demands real-time semantic situation analysis within milliseconds, particularly for resource-constrained edge devices. Limited data collection windows (seconds to minutes) and statistical non-stationarity of signal characteristics violate assumptions of traditional machine learning models, complicating fast and robust semantic extraction.

To meet the real-time requirements in dynamic environments, this subsection introduces innovative learning techniques specifically designed for fast semantic situation analysis. The possible approaches include edge-adaptive neural network architectures that dynamically adjust structures based on signal complexity and available computational resources, making real-time semantic analysis possible even on resource-constrained devices \cite{zhang2023lightweight}; context-aware few-shot learning system enhanced with contextual information to recognize new signal types with minimal samples \cite{zhang2024sswsrnet}, enabling quick adaptation to emerging communication protocols; and distributed continual learning \cite{zhu2020toward} ecosystem implementing knowledge distillation and transfer mechanisms that enable collaborative learning across edge devices while preserving privacy, creating a network-wide semantic understanding that continuously improves without requiring complete retraining.

\subsection{Security Vulnerabilities in Open Wireless Networks}

Open wireless networks expose AI-based spectrum cognition models to adversarial attacks \cite{kim2021channel,wang2023adversarial}, where minimal signal perturbations can cause misclassification, compromising semantic situation analysis. Distinguishing natural signal variations from malicious manipulations is increasingly difficult, and robust defenses often degrade accuracy or speed, creating a trade-off between security and efficiency. This is particularly critical in 6G, where semantic analysis informs autonomous resource allocation and security measures.

To address security vulnerabilities in AI-based semantic situation analysis, this subsection presents innovative defense mechanisms that maintain security without significantly compromising performance. The proposed defenses include multi-level signal authentication frameworks with three protection layers (physical characteristic analysis, statistical anomaly detection, and semantic consistency verification) to ensure trustworthy semantic information extraction; adaptive feature obfuscation mechanisms employing dynamic feature transformations to maintain signal identifiability while preventing attackers from determining attack targets; and integrated heterogeneous model collaborative defense combining physics-based and learning-based models \cite{bakirtzis2025empowering} with novel voting mechanisms to maintain system accuracy even under adversarial conditions, ensuring that no single attack can compromise the entire semantic situation analysis system.

\section{Conclusion} 
\label{sec_conclusion}

This article presents spectrum cognition as a transformative approach for next-generation wireless communications, introducing semantic situation as the core concept for intelligent spectrum understanding beyond traditional signal detection. The proposed three-layer semantic hierarchy enables comprehensive spectrum analysis from basic device identification to communication purpose inference, supporting diverse 6G applications while addressing challenges in complex electromagnetic environments, real-time processing, and security vulnerabilities. Future research should focus on scalability, cross-domain cognition, and practical deployment considerations. Spectrum cognition with semantic situation awareness represents a critical enabling technology for 6G networks, providing intelligent spectrum understanding necessary for efficient, secure, and user-centric wireless communications.


\bibliographystyle{IEEEtran}
\bibliography{IEEEabrv,mag_ref_s}

\vspace{-10 mm}

\begin{IEEEbiographynophoto}{Hao Zhang}
received his Ph.D. degree from Nanjing University of Aeronautics and Astronautics, China, in 2025. He is now a Post-Doctoral Fellow at the College of Artificial Intelligence, Nanjing University of Aeronautics and Astronautics. His current research interests include foundation models, spectrum cognition, and wireless communication systems.
\end{IEEEbiographynophoto}

\vspace{-15 mm} 

\begin{IEEEbiographynophoto}{Fuhui Zhou} is a professor with the College of Artificial Intelligence, Nanjing University of Aeronautics and Astronautics.  His current research interests include cognitive radio, RF machine learning, and knowledge graphs.
\end{IEEEbiographynophoto}

\vspace{-15 mm} 

\begin{IEEEbiographynophoto}{Qihui Wu} is a professor with the College of Electronic and Information Engineering, Nanjing University of Aeronautics and Astronautics. His current research interests span the areas of wireless communications and statistical signal processing.
\end{IEEEbiographynophoto}

\vspace{-15 mm}

\begin{IEEEbiographynophoto}{Chau Yuen} is an associate professor at the School of Electrical \& Electronic Engineering, Nanyang Technological University. His current research interests include Holographic MIMO Surface and Reconfigurable Intelligence Surface. 
\end{IEEEbiographynophoto}

\end{document}